\documentclass [review,authoryear]{elsarticle}

\usepackage[version=4]{mhchem}

\usepackage[utf8]{inputenc}
\usepackage{multirow}
\usepackage[table,xcdraw]{xcolor}
\usepackage{graphicx}
\begin{document}
\title{Simple photochemical modelling of $\ce{NO_X}$ pollution in a street canyon}
\author[1]{L. Soulhac}
\author[1]{S. Fellini}
\author[2]{C.V. Nguyen}
\author[2]{P. Salizzoni}
\address[1]{Univ. Lyon, INSA Lyon, CNRS, Ecole Centrale de Lyon, Univ. Claude Bernard Lyon 1, LMFA, UMR5509, 69621, Villeurbanne France}
\address[2]{Univ. Lyon, Ecole Centrale de Lyon, CNRS, Univ. Claude Bernard Lyon 1, INSA Lyon, LMFA, UMR5509, 69130, Ecully, France}
\maketitle

\section*{Abstract}
To predict pollutant concentration in urban areas, it is crucial to take into account the chemical transformations of reactive pollutants in operational dispersion models. In this work, we derive and discuss different \ce{NO-NO_2-O_3} chemical street canyon models with increasing complexity and we analytically evaluate their applicability in different urban contexts. We then evaluate the performance of the models in predicting \ce{NO_2} concentration at different locations within an urban district by comparing their predictions with measurements acquired in a field campaign.
The results are in line with analytical speculations and give indications as to which model to use according to the conditions of the urban street canyon. In courtyards with limited ventilation and without direct emissions, the performance of the photostationary model is satisfactory. On the other hand, the application of a non-photostationary model significantly improves the predictions in urban canyons with direct vehicular emissions. The applicability of the proposed models in operational tools at the city scale is finally discussed. 

\section*{Keywords}
Photochemical smog, Urban air quality, Non-photostationary chemical model, Air pollution measurement campaign

\section*{Introduction}

The time scales related to pollutant transfer over large urban agglomerations range from a few minutes to several hours. During this period, a large number of physico-chemical processes take place and determine the concentration of pollutants in the urban atmosphere \citep{sillman1999relation}.
When the focus is on dispersion at the local district scale, the rate of turbulent transport is considerably high compared to the rate of chemical transformation and most of the  atmospheric compounds can be treated as inert tracers. There are however chemical reactions which are sufficiently fast to significantly affect the concentration of pollutants during their residence time in the streets. This is notably the case for $\ce{NO_X}$, i.e. the nitrogen oxides that are most relevant for air pollution. 

The emissions of $\ce{NO_X}$ result from combustion processes, especially from motor vehicle engines or from power stations and industries. They are therefore a tracer of anthropogenic activity in urban areas and their trends are used to assess the effectiveness of regulations on air pollution, or to evaluate the effects of sudden changes in emissions, such as during COVID-19 restrictions \citep[e.g.,][]{toscano2020effect,lovarelli2020describing,misra2021nitrogen}. It is generally assumed that the partition of $\ce{NO_X}$ at the point of emission is approximately between 10\% to 15\% for $\ce{NO_2}$ (nitrogen dioxide) and 85\% to 90\% for $\ce{NO}$ (nitrogen monoxide) \citep{ntziachristos2000copert}.
Acute exposure to $\ce{NO_X}$ causes respiratory disease and compromises lung functioning when inhaled at high concentrations. Children are the most vulnerable, with a demonstrated increased incidence of childhood asthma due to $\ce{NO_2}$ emissions from vehicular traffic \citep{khreis2017exposure,anenberg2022long}. Despite being the major contributor to $\ce{NO_X}$, $\ce{NO}$ is less toxic than $\ce{NO_2}$. However, as most radicals, it is extremely unstable and forms $\ce{NO_2}$ through photochemical oxidation. Nitrogen dioxide is then converted back to $\ce{NO}$ as a result of photolysis which also leads to the regeneration of ozone ($\ce{O_3}$). 
When the photostationary state is reached, these reactions result in a cycle with zero net chemistry and the chemical compounds reach the equilibrium composition, which can be easily derived in terms of kinetic reaction parameters by the Leighton relation \citep{leighton2012photochemistry}. 
Deviations from this state occur when (i) the residence time of polluants in the reference volume (i.e. the street) is shorter than the time needed for reaching the photostationary equilibrium, (ii) turbulent motions mix the reactants so slowly that they remain segregated rather than reacting \citep{li2021error}, (iii) the transformation of nitrogen monoxide into $\ce{NO_2}$ is altered by the role of complex reactions with radicals resulting from the oxidation of Volatile Organic Compounds ($\ce{VOC}$s) and $\ce{CO}$ \citep{jenkin2000ozone}. The concentrations of $\ce{NO}$ and $\ce{NO_2}$ are also affected by reactions involving the hydroxyl radical and leading to the production of nitric acid.

The coupling of turbulent and chemical dynamics to assess photochemical pollution in urban areas has been explored extensively in the past two decades by means of Computational Fluid Dynamics (CFD) simulations. \citet{baker2004study} extended a Large Eddy Simulation (LES) for turbulent flow in a street canyon with a simple $\ce{NO_X}$-$\ce{O_3}$ chemical model. The same reaction scheme was adopted by \citet{baik2007modeling}, who instead used Reynolds-averaged Navier–Stokes (RANS) simulations. By introducing a photostationary state defect index, both studies highlighted the regions of a street canyon most prone to chemical instability. The chemistry of $\ce{VOC}$ has been included in RANS simulations by \citet{kwak2012cfd} and \citet{kim2012urban}, while \citet{bright2013urban} combined LES simulations with a detailed chemical reaction mechanism (Reduced Chemical Scheme) comprising 51 chemical species and 136 reactions. Similarly, \citet{garmory2009simulations} used the Stochastic Fields (FS) method to simulate turbulent reacting flows with a chemistry model comprising 28 species. 
These studies showed that the effect of turbulent fluctuations (i.e. segregation) on the chemistry is significant for species with the highest transformation rates. They also showed that increasing chemical complexity (i.e. simulating $\ce{VOC}$ chemistry) could contribute to additional but modest $\ce{NO_2}$ and $\ce{O_3}$ formation in the canyon. 

CFD simulations, coupled with detailed chemical models, provide an accurate prediction but are computationally expensive and require a large amount of detailed input data. To simulate air quality in large urban domains, consisting of hundreds to thousands of streets, a more efficient way is adopting simplified modelling approaches \citep{vardoulakis2007operational}.
These are usually Gaussian-Lagrangian models integrated with box models to simulate the concentration in the street canyons.
In these operational tools, photostationarity is a convenient assumption as it allows the modelling of $\ce{O3}$ and $\ce{NO_X}$ as inert tracers and to subsequently apply photochemical equilibrium in the streets. This is the case of the Canyon Plume Box Model (CPBM) \citep{yamartino1986development}, and the street network model Sirane \citep{soulhac2017model}.

Another widespread approach is the adoption of empirical models to estimate $\ce{NO}$-$\ce{NO_2}$ conversion \citep{ravina2022air}. These are based on a photostationary assumption but are optimised to fit observed concentrations. 
\citet{hirtl2007evaluation} investigated the performances of the two empirical conversion schemes after \citet{romberg1996n0} and after \citet{derwent1996empirical}, when implemented in the Gaussian model Atmospheric Dispersion Modelling System (ADMS) and in the LAgrangian Simulation of Aerosol Transport (LASAT) model. These dispersion models turned out to be quite successful in predicting average concentrations measured in street canyons.

A step forward in modeling the interaction between the time scales of chemical reactions and those of transport is represented by the model ADMS-Urban \citep{mchugh1997adms,carruthers2000use}. 
In ADMS-Urban, $\ce{NO_X}$ chemistry can be modelled by the Generic Reaction Set (GRS) \citep{azzi1992introduction,venkatram1994development} photochemical scheme which includes seven chemical reactions. 
The GRS chemical model is applied to the emitted pollutants after transport and dispersion. The chemistry calculation for the receptor is split in two steps: the first considers the contribution from far sources (source-receptor travel time greater than 150 s), while the second one includes the contribution from the nearest sources (source-receptor travel time less than 150 s) \citep{ADMS_tech}. In this way, the model takes into account the travel time of the pollution plume and it assumes a time -or distance- dependence on the generation of $\ce{NO_2}$.

Finally, the Operational Street Pollution Model (OSPM) \citep{palmgren1996effects,berkowicz1997modelling} is a street canyon model which includes \ce{NO}-\ce{NO_2}-\ce{O_3} chemistry by means of a non-photostationary model that takes into account the interaction between the chemical reaction rates and the residence time of the pollutants in the street.

The overview above suggests that operational modeling of reactive pollutant concentration at the urban scale requires an adequate description of (i) the chemistry, (ii) the turbulent transport, (iii) the interaction between these two processes, all while minimizing the computational cost and required input data in order to be applied to hundreds to thousands of streets. 
To date, empirical relationships and photostationary models are the most commonly used for operational purposes while non-photostationary schemes are rarely implemented. This is especially true for street network models, where, to our knowledge, the non-photostationary scheme has not yet been implemented.
Furthermore, the existing literature lacks a coherent formulation of the different photochemical models, with a clear statement of the underlying assumptions and a concurrent validation with real data. 
To fill these gaps, in this work, we derive, compare and validate three models for $\ce{NO_X}$ photochemical pollution that can be efficiently implemented in street network models at the city scale.  
In the analytical derivation, we focus on the time scales of pollutant transformation and transport in order to highlight the range of application of the different models.  
To verify the reliability of the different schemes we compare the model outputs to field data. The main objective is to evaluate whether the application of a non-photostationary model can bring substantial advantages in the prediction of pollutant concentration in the streets, with respect to photostationary models.

The formulation of a photochemical model, adopting box-model approach, is presented in \ref{section_model}.
A general presentation of the measurement campaign is given in Section \ref{measurements}. Results are discussed in Section \ref{results}, while the conclusions are drawn in Section \ref{conclusions}.

\section{$\ce{NO}-\ce{NO_2}-\ce{O_3}$ chemical street model}
\label{section_model}

To maximize computational efficiency, minimize input data while providing a satisfactory description of pollution in the urban area, city-scale operational models, such as street network models \citep{soulhac2011model}, generally provide a single concentration value for each street. This can be notably achieved adopting a bow model a the street scale, which provides spatially averaged pollutant concentration by computing a pollutant budget over the volume of the street.
In order to simulate photochemical pollution, the pollutant budget has to take into account the terms of chemical production and of chemical destruction \citep{soulhac2011model} as well as those related to the turbulent fluxes at the street edges and at the top of the street.\\

To write the budget of photochemical pollutants in the street, we start by considering the simplified chemical scheme involving \ce{NO}, \ce{NO_2} and \ce{O_3}:

\begin{equation}
\label{Equ-Mreac1}
\ce{NO_2} \stackrel{k_1}{\longrightarrow}\ce{NO}+\ce{O^\bullet}
\end{equation}

\begin{equation}
\label{Equ-Mreac2}
\ce{O_2}+\ce{O^\bullet} \stackrel{k_2}{\longrightarrow}\ce{O_3}
\end{equation}

\begin{equation}
\label{Equ-Mreac3}
\ce{NO}+\ce{O_3} \stackrel{k_3}{\longrightarrow}\ce{NO_2}+\ce{O_2}
\end{equation}

It is known \citep{seinfeld1986atmospheric} that the second equation is much faster than the first and the third ones, so that the constants $k_1$ and $k_3$ are the limiting parameters of these chemical reactions.
The constant rate $k_1$ ($\ce{NO_2}$ photolysis rate) depends on the intensity of solar radiation, whilst $k_3$ depends on air temperature. These dependences can be modeled by the following relations \citep{kasten1980solar,seinfeld1986atmospheric}: 

\begin{equation}\label{k1k3}
\left\{
\begin{array}{l}
k_1=\frac{1}{60}(0.5699-[9.056\cdot 10^{-3}(90-\zeta)]^{2.546})\left(1-0.75\left[\frac{Cld}{8}\right]^{3.4}\right) \, (\mbox{s}^{-1})\\
k_3=1.325\cdot 10^{6}\exp\left(-\frac{1430}{T}\right)   \, (\mbox{m}^3 \mbox{mol}^{-1}\mbox{s}^{-1})
\end{array}
\right.
\end{equation}
where $\zeta$ is the solar elevation in degrees, $T$ is the air temperature in Kelvin and $Cld$ is the cloud coverage in Oktas. These meteorological parameters vary over time. In operational dispersion models, the time-dependence of the meteorological parameter is usually modelled assuming a quasi-steady approach, i.e. assuming steady condition of time step of 1 hours. Cloud coverage and temperature are measured during the day at meteorological stations, while the solar elevation is a function of the day of the year, the local hour and the site latitude \citep[e.g.,][]{soulhac2011model}. Note that $k_1$ is set equal to 0 at night, when the solar elevation angle is negative.  
More sophisticated models for $k_1$ and $k_3$ are available in the literature, but they are generally not adapted for operational purposes \citep{seinfeld1986atmospheric}. 

Referring to Eqs. \ref{Equ-Mreac1}-\ref{Equ-Mreac3}, the production and destruction terms for each chemical species are related to the molar concentration by the following expressions:

\begin{equation}\label{prod_destr}
    \begin{array}{ll}
        P_{NO} =k_1 [\ce{NO_2}] & D_{NO}=k_3 [\ce{NO}][\ce{O_3}]  \\
        P_{NO_2} =k_3 [\ce{NO}][\ce{O_3}] & D_{NO_2}=k_1 [\ce{NO_2}]  \\
        P_{O_3} =k_1 [\ce{NO_2}] & D_{O_3}=k_3 [\ce{NO}][\ce{O_3}]  \\
         & 
    \end{array}
\end{equation}
where $[\ce{\cdot}]$ represents the molar concentration (mol/$\mbox{m}^3$) of the compound. \\

 We include the production and destruction terms in the street box model formulated in \citet{soulhac2011model}.  
 Neglecting wet and dry deposition phenomena, the budget of time-averaged concentration of \ce{NO_2}, \ce{NO} and \ce{O_3} for a single street-canyon of length $L$, width $W$ and height $H$, can be written:

\begin{equation}
\label{Equ-Adv-Diff}
Q_{\ce{NO_2}} - u_d \mathcal{S}_h \left( [\ce{NO_2}] - [\ce{NO_2}]\!^r \right) - U \mathcal{S}_v \left( [\ce{NO_2}] - [\ce{NO_2}]\!^c \right) + k_3 [\ce{NO}] [\ce{O_3}] \mathcal{V} - k_1 [\ce{NO_2}] \mathcal{V} = 0
\end{equation}
\begin{equation}
\label{Equ-Adv-Diff_NO}
Q_{\ce{NO}} - u_d \mathcal{S}_h \left( [\ce{NO}] - [\ce{NO}]\!^r \right) - U \mathcal{S}_v \left( [\ce{NO}] - [\ce{NO}]\!^c \right) + k_1 [\ce{NO_2}] \mathcal{V} - k_3 [\ce{NO}][\ce{O_3}] \mathcal{V} = 0
\end{equation}
\begin{equation}
\label{Equ-Adv-Diff_O3}
- u_d \mathcal{S}_h \left( [\ce{O_3}] - [\ce{O_3}]\!^r \right) - U \mathcal{S}_v \left( [\ce{O_3}] - [\ce{O_3}]\!^c \right) + + k_1 [\ce{NO_2}] \mathcal{V} - k_3 [\ce{NO}][\ce{O_3}] \mathcal{V} = 0
\end{equation}
where $\mathcal{V} = L W H$ is the volume of the street, $\mathcal{S}_h = L W$ is its horizontal area and $\mathcal{S}_v = W H$ is its vertical cross section. 
The velocities $U$ and $u_d$ are the mean velocity along the street and the exchange rate at roof level \citep[e.g.,][]{soulhac2008flow,salizzoni2009street,fellini2020street} and they drive the longitudinal and vertical pollutant fluxes entering and leaving the street volume.
For each of the three chemical compounds \ce{NO_2}, \ce{NO} and \ce{O_3}, $Q$ is the molar emission in the street, $[\cdot]\!^r$ is the concentration in the atmosphere above roofs, and $[\cdot]\!^c$ is the concentration in the flow advected within the canopy at the upwind intersection of the street. 
We point out that the source of ozone is not included in the budget (i.e. $Q_{\ce{O_3}}=0$) since direct ozone emissions in the streets are rare. 
Eqs. \ref{Equ-Adv-Diff}-\ref{Equ-Adv-Diff_O3} can be reformulated by highlighting the time scales associated with the terms of transport and chemical reaction. For example, for $\ce{NO_2}$ we can write:
\begin{equation}
\label{Equ-Adv-Diff_times}
\dfrac{Q_{\ce{NO_2}}}{\mathcal{V}} - \dfrac{[\ce{NO_2}] - [\ce{NO_2}]\!^r}{\tau_v} - \dfrac{[\ce{NO_2}] - [\ce{NO_2}]\!^c}{\tau_h} + \dfrac{[\ce{NO}]}{\tau_3}  - \dfrac{ [\ce{NO_2}]}{\tau_1} = 0
\end{equation}

\noindent with
\begin{equation}
\left\{
\begin{array}{l}
\tau_v = \dfrac{H}{u_d}\\
\tau_h = \dfrac{L}{U}\\
\tau_1 = \dfrac{1}{k_1}\\
\tau_3 = \dfrac{1}{k_3[\ce{O_3}]}
\end{array}
\right.
\end{equation}

\noindent To determine the order of magnitude of the different terms in Eq. \ref{Equ-Adv-Diff_times} we can roughly estimate the time scales involved, based on the data collected and simulated for the city of Lyon (France) \citep{soulhac2010dispersion,soulhac2012model}. The depth $H$ and length $L$ of street canyons vary in the ranges 15-30 m and 20-150 m, respectively. The wind speed within the streets $U$ and the typical turbulent exchange velocity $u_d$ can reasonably be assumed in the ranges 0.1-5 m/s and 0.01-0.22 m/s, respectively \citep{salizzoni2009street,soulhac2011model} when the free stream wind above the city is between 1.5 m/s and 8 m/s (Météo-France data for the period 1981–2006)
From these data, we obtain that $\tau_v$ ranges in 68-3000 s, and $\tau_h$ in 4-1500 s. Typical values of $k_1$ and $k_3$ can be estimated by means of Eq. \ref{k1k3} by varying the cloud coverage $Cld$ between 0 and 8, the temperature $T$ in $5^o$C-$30^o$C and $\zeta$ in $10^o$-$90^o$ (in this analysis we consider only daytime). The concentration of ozone can be taken in the range 25-75 ppb (data measured at Saint-Exupery station for the year 2008). 
These data provide $\tau_1$ in the range 105-1850 s and $\tau_3$ in the range 54-247 s.
Moreover, we introduce an average time scale $\tau_s$ related to the pollutant wash-out from the street:
\begin{equation}
\label{Equ-taus}
\tau_s = \left( \dfrac{1}{\tau_h} + \dfrac{1}{\tau_v} \right)^{-1}
\end{equation}
and we find that $\tau_s$ varies approximately in the range 4-1000 s. This analysis shows that there is an overlap between the timescales associated to chemical reactions and the characteristic residence times of pollutants within the street. Consequently a modeling approach combining chemistry and advection-diffusion processes must be adopted, as neither of the two processes can be neglected.\\ 

Finally, we define the average background concentration $[\cdot]\!^b$ as:
\begin{equation}
[\cdot]\!^b = \dfrac{\dfrac{[\cdot]\!^c}{\tau_h} + \dfrac{[\cdot]\!^r}{\tau_v}}{\dfrac{1}{\tau_h} + \dfrac{1}{\tau_v}}
\end{equation}.

\noindent In this way, Eq. \ref{Equ-Adv-Diff_times} can be simplified using only the background concentration $[\ce{NO_2}]\!^b$:
\begin{equation}
\label{Equ-Adv-Diff_final}
\dfrac{Q_{\ce{NO_2}}}{\mathcal{V}} - \dfrac{[\ce{NO_2}] - [\ce{NO_2}]\!^b}{\tau_s} + k_3 [\ce{NO}] [\ce{O_3}] - k_1 [\ce{NO_2}] = 0
\end{equation}
The same formulation is valid for \ce{NO} and \ce{O_3} so that a system of 3 equations (Eq. \ref{Equ-Adv-Diff_final} and the two analogous balances for \ce{NO} and \ce{O_3}) describes in a compact way the dynamics of the three chemical compounds. 
In what follows, we will examine the solution of this system of equations adopting different scenarios related to the relative importance of the different time scales involved.

\subsection{Passive scenario}
\label{sec:Passive-street-Theory}

As a first step, we consider the case of a passive pollutant, whose concentration is generally referred to as $[\cdot]\!^*$. 
For \ce{NO_2} Eq. \ref{Equ-Adv-Diff_final} simplifies as:
\begin{equation}
\label{Equ-Adv-Diff-Passive}
\dfrac{Q_{\ce{NO_2}}}{\mathcal{V}} - \dfrac{[\ce{NO_2}]\!^* - [\ce{NO_2}]\!^b}{\tau_s} = 0.
\end{equation}
This scenario corresponds to the case of reaction times that are extremely long (i.e. $\tau_1\rightarrow \infty$, $\tau_3\rightarrow \infty$) so that the terms of chemical production and destruction are negligible for the budget in the street.
\noindent The solution is given by:
\begin{equation}
\label{Equ-Passive-NO2}
[\ce{NO_2}]\!^* = [\ce{NO_2}]\!^b + \dfrac{\tau_s Q_{\ce{NO_2}}}{\mathcal{V}}
\end{equation}

By analogy, the relative solution for the `passive' \ce{NO} concentration reads:
\begin{equation}
\label{Equ-Passive-NO}
[\ce{NO}]\!^* = [\ce{NO}]\!^b + \dfrac{\tau_s Q_{\ce{NO}}}{\mathcal{V}}
\end{equation}

\noindent and for \ce{O_3} concentration, assuming no emission of ozone:
\begin{equation}
\label{Equ-Passive-O3}
[\ce{O_3}]\!^* = [\ce{O_3}]\!^b.
\end{equation}

The concentration $[\cdot]\!^*$ takes into account all contributions to pollution deriving from advective transport only, i.e. the direct emission into the street and the transport of pollutants to the street both from adjacent street and from the atmosphere above the roofs.
Thus, Eqs. \ref{Equ-Passive-NO2}-\ref{Equ-Passive-O3} can be seen as the general solutions for a dispersion model able to provide passively advected concentrations in the streets. 

%================================================================
\subsection{Photostationary chemical model}
\label{sec:Photostationary-Theory}

Let us now consider that the reactive pollutants in the control volume (i.e. within the street canyon) have the necessary time to reach the photochemical equilibrium.
This corresponds to assume that the characteristic time scales of the chemical reactions $\tau_1$ and $\tau_3$ are small compared to the residence time of pollutants within the street (i.e. $\tau_1$ and $\tau_3 \longrightarrow 0$).
Under these assumptions, the advective and source terms in Eq. \ref{Equ-Adv-Diff_final} become negligible compared to the production and destruction terms and the balance equation is simplified as follows:  

\begin{equation}
\label{Equ-Photostationary-state}
k_3 [\ce{NO}]\!^{\infty} [\ce{O_3}]\!^{\infty} - k_1 [\ce{NO_2}]\!^{\infty} = 0
\end{equation}

\noindent where the photostationary concentrations have been referred to as $[\cdot]\!^{\infty}$. Eq. \ref{Equ-Photostationary-state} is known as the Leighton relationship \citep{leighton2012photochemistry}, whose formulation could also be obtained from the budget of \ce{NO_2} or \ce{O_3} (see Eqs. \ref{Equ-Adv-Diff_NO} and \ref{Equ-Adv-Diff_O3}). 

The conservation of \ce{N} and \ce{O} species lead to the following relations, which are valid for passive, photostationary or non-photostationary concentrations:
\begin{equation}
\label{Equ-Stoechiometry-N}
[\ce{NO}] + [\ce{NO_2}] = [\ce{NO}]\!^* + [\ce{NO_2}]\!^* = [\ce{NO}]\!^{\infty} + [\ce{NO_2}]\!^{\infty} = \phi_N,
\end{equation}

\begin{equation}
\label{Equ-Stoechiometry-O}
[\ce{O_3}] + [\ce{NO_2}] = [\ce{O_3}]\!^* + [\ce{NO_2}]\!^* = [\ce{O_3}]\!^{\infty} + [\ce{NO_2}]\!^{\infty} = \phi_O,
\end{equation}

\noindent where $\phi_N$ and $\phi_O$ are constants defining the proportion of the different species, whatever the chemical history of the pollutants reaching the street canyon. We note that $\phi_N$ and $\phi_O$ can be easily computed from the results of the passive model providing the concentrations $[\cdot]^*$ (Section \ref{sec:Passive-street-Theory}) which take into account all the pollutant contributions reaching the canyon (i.e. both direct emissions and transported pollutants).

Combining Eqs. \ref{Equ-Photostationary-state} to \ref{Equ-Stoechiometry-O} (see e.g.,  \citet{soulhac2011model}) provides the solution:

\begin{equation}
\label{Equ-Photostationary-Model}
[\ce{NO_2}]\!^{\infty} = \dfrac{b - \sqrt{b^2 - 4 c}}{2}
\end{equation}

\noindent with:
\begin{equation}
\label{Equ-Photostationary-Csts}
\left\{
\begin{array}{l}
b = \dfrac{k_1}{k_3} + [\ce{O_3}]\!^* + [\ce{NO}]\!^* + 2 [\ce{NO_2}]\!^* = \dfrac{k_1}{k_3} + \phi_N + \phi_O\\[0.4cm]
c = \left( [\ce{O_3}]\!^* + [\ce{NO_2}]\!^* \right) \left( [\ce{NO}]\!^* + [\ce{NO_2}]\!^* \right) = \phi_O . \phi_N
\end{array}
\right.
\end{equation}

Eq. \ref{Equ-Photostationary-Csts} illustrates that \ce{NO_2} concentration depends only on $k_1 / k_3$, $\phi_N$ and $\phi_O$. This highlights that the chemical history of the background concentration, that is included in passive concentrations $[\cdot]\!^*$ by equations \ref{Equ-Passive-NO2} to \ref{Equ-Passive-O3}, has no influence on the photostationary solution because this solution corresponds to an infinite reaction time, which offsets the initial repartition between \ce{NO}, \ce{NO_2} and \ce{O_3}.

Once $[\ce{NO_2}]\!^{\infty}$ is known, Eqs. \ref{Equ-Stoechiometry-N} and \ref{Equ-Stoechiometry-O} provide $[\ce{NO}]\!^{\infty}$ and $[\ce{O_3}]\!^{\infty}$.

%================================================================
\subsection{Non-photostationary chemical model}
\label{sec:Chem-street-Theory}

To find the general solution for the full chemical street model (Eq.\ref{Equ-Adv-Diff_final}), we take the difference between Eqs. \ref{Equ-Adv-Diff_final} and \ref{Equ-Adv-Diff-Passive}:

\begin{equation}\label{21}
- \dfrac{[\ce{NO_2}] - [\ce{NO_2}]\!^*}{\tau_s} + k_3 [\ce{NO}] [\ce{O_3}] - k_1 [\ce{NO_2}] = 0
\end{equation}

By introducing Eqs. \ref{Equ-Stoechiometry-N} and \ref{Equ-Stoechiometry-O} in Eq. \ref{21}:
\begin{equation}
\label{Equ-2nd_degree}
- \dfrac{[\ce{NO_2}] - [\ce{NO_2}]\!^*}{\tau_s} + k_3 \left( [\ce{NO}]\!^* + [\ce{NO_2}]\!^* - [\ce{NO_2}] \right) \left( [\ce{O_3}]\!^* + [\ce{NO_2}]\!^* - [\ce{NO_2}] \right) - k_1 [\ce{NO_2}] = 0
\end{equation}

\noindent and rearranging, the equation for $[\ce{NO_2}]$ is finally:
\begin{equation}
[\ce{NO_2}]^2 - b^\prime [\ce{NO_2}] + c^\prime = 0
\end{equation}

\noindent with
\begin{equation}
\label{Equ-NonPhotostationary-Csts}
\left\{
\begin{array}{l}
b^\prime = b + \dfrac{1}{k_3 \tau_s}\\[0.4cm]
c^\prime = c + \dfrac{[\ce{NO_2}]\!^*}{k_3 \tau_s}
\end{array}
\right.
\end{equation}

We obtain the non-photostationary solution for $[\ce{NO_2}]$ is then:
\begin{equation}
\label{Equ-NonPhotostationary-Model}
[\ce{NO_2}] = \dfrac{b^\prime - \sqrt{{b^\prime}^2 - 4 c^\prime}}{2}
\end{equation}
This expression is very similar to the photo-chemical model implemented in OSPM \citep{palmgren1996effects,berkowicz1997modelling} but generalized for a street canyon with a longitudinal advection velocity $U$ and a vertical turbulent exchange rate $u_d$, therefore suitable for implementation in street network models. 
Also in this case, once $[\ce{NO_2}]$ is known, Eqs. \ref{Equ-Stoechiometry-N} and \ref{Equ-Stoechiometry-O} provide $[\ce{NO}]$ and $[\ce{O_3}]$.

\begin{figure}
    \centering
    \includegraphics[width=12cm]{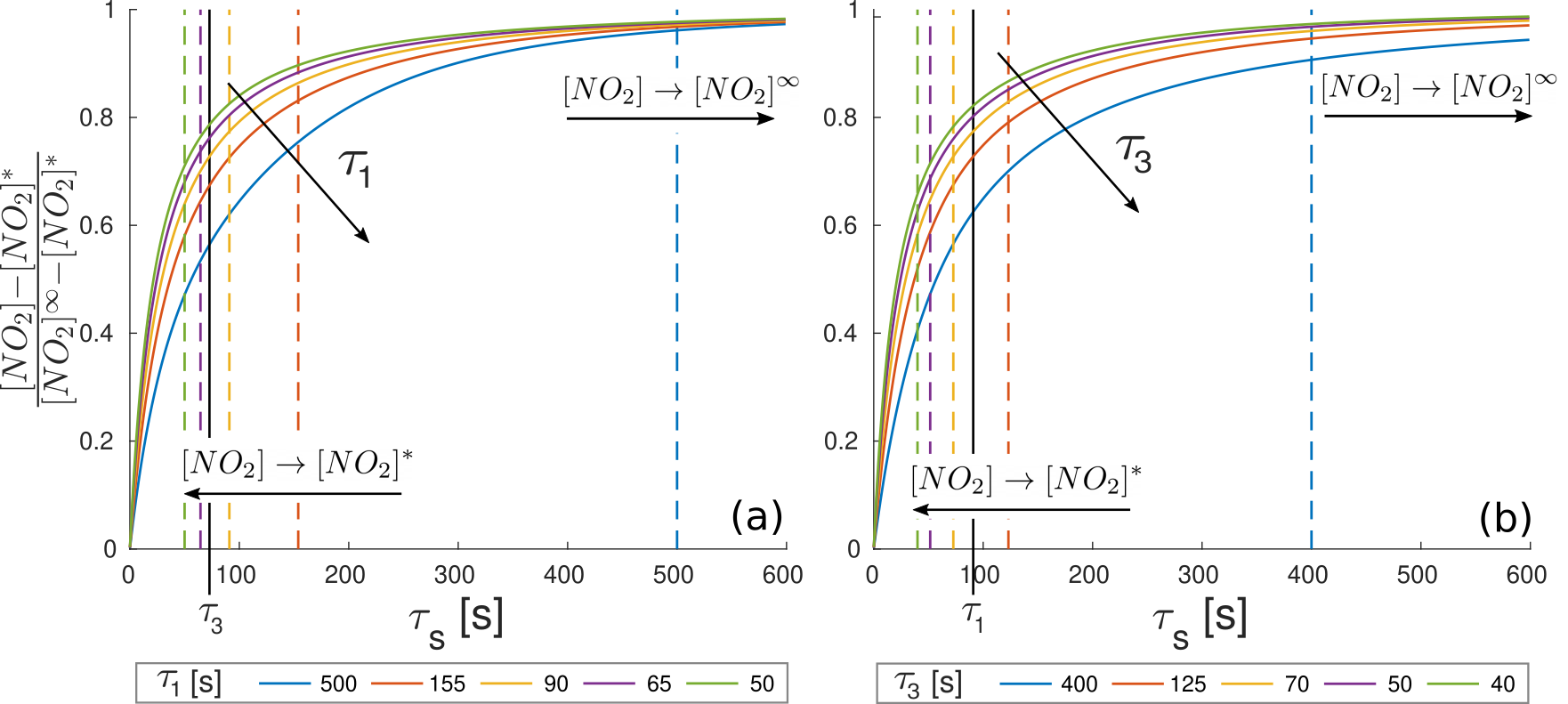}
    \caption{Trend of the non-photostationary solution ($[\ce{NO_2}]$) towards the solution for the passive model ($[\ce{NO_2}]^*$) and towards the photostationary solution ($[\ce{NO_2}]^\infty$) as a function of the characteristic time of advective transport ($\tau_s$), and of the two characteristic times of reaction. Panel (a) shows different curves as a function of $\tau_1$, indicated in the legend and by the dashed vertical lines. Similarly, panel (b) shows the curves as a function of $\tau_3$ indicated in the legend and by the dashed vertical lines.}
    \label{fig:trend_tau_tau3}
\end{figure}

The solution in Eqs. \ref{Equ-NonPhotostationary-Csts}-\ref{Equ-NonPhotostationary-Model} can be discussed according to the asymptotic values for the street residence time scale $\tau_s$, as shown in Fig. \ref{fig:trend_tau_tau3}. 
If $\tau_s \rightarrow 0$ (i.e. $\tau_s \ll \tau_1$ and $\tau_3$ ), according to Eq. \ref{Equ-2nd_degree}, we have that $[\ce{NO_2}] \rightarrow [\ce{NO_2}]\!^*$. 
It means that the pollutants have no time to react and the concentration of the chemical species is provided by the passive solution (Eqs. \ref{Equ-Passive-NO2}-\ref{Equ-Passive-O3}). On the other hand, if $\tau_s$ tends to infinity (i.e. $\tau_s \gg \tau_1$ and $\tau_3$ ), then $b=b'$ and $c=c'$ in Eq. \ref{Equ-NonPhotostationary-Csts} and the concentration $[\ce{NO_2}]$ tends to $[\ce{NO_2}]\!^{\infty}$. It means that the pollutants have an infinite time to react and the final concentration of the chemical species is provided by the photostationary solution (Eq. \ref{Equ-Photostationary-Model}).

For intermediate values of $\tau_s$, the solution is in-between, with only a partial conversion from \ce{NO} to \ce{NO_2}, compared to the photostationary limit. We can remark that, unlike the photostationary case, Eq. \ref{Equ-NonPhotostationary-Csts} includes $[\ce{NO_2}]\!^*$ independently of the constants $\phi_N$ and $\phi_O$. This adds a dependence of the solution on the chemical history of the background concentration.\\

To provide further interpretations, Eq. \ref{Equ-2nd_degree} can be rewritten in the form:

\begin{equation}
\underbrace{[\ce{NO_2}]\!^*}_{\alpha} \chi^2 - \underbrace{\left( \dfrac{k_1}{k_3} + [\ce{O_3}]\!^* + [\ce{NO}]\!^* + \dfrac{1}{k_3 \tau_s} \right)}_{\beta} \chi + \underbrace{\left( \dfrac{[\ce{NO}]\!^* [\ce{O_3}]\!^*}{[\ce{NO_2}]\!^*} - \dfrac{k_1}{k_3} \right)}_{\gamma} = 0
\end{equation}

\noindent with
\begin{equation}
\chi = \dfrac{[\ce{NO_2}] - [\ce{NO_2}]\!^*}{[\ce{NO_2}]\!^*} = \dfrac{\beta - \sqrt{\beta^2 - 4 \alpha \gamma}}{2 \alpha}.
\end{equation}

\noindent where $\chi$ represents the rate of increase of \ce{NO_2} concentration with respect to its passive value $[\ce{NO_2}]\!^*$. The term $\gamma$ can be seen as a quantification of the non-photostationarity of the passive concentrations of \ce{NO}, \ce{NO_2} and \ce{O_3}. Consequently, if the passive concentrations are already close to the photostationary equilibrium ($\gamma \simeq 0$), then the solution tends to $\chi \simeq 0$, which means that the final concentration is close to the passive one.\\

\subsection{Validation strategy}\label{three_models}
As a further step, we test the analytical solutions presented in the previous sections against field data. In doing so, we will consider three different models: the first one (Model 1) is the photostationary model (Eq. \ref{Equ-Photostationary-Model}) where the transformation rates $k_1$ and $k_3$ are assumed to be constant with time, whatever the temperature and radiative conditions. This solution can be adopted when the meteorological information (temperature, intensity of solar radiation) is missing and to minimize the computational cost. 
The sensitivity of the model to the value adopted for the $k_1/k_3$ ratio is discussed in the following section.  
The second one (Model 2) is a photostationary model with transformation rates that vary during the day according to the meteorological conditions. To this aim, Eq. \ref{k1k3} is applied to estimate $k_1$ and $k_3$ with the parameters $T$, $Cld$ and $\zeta$ varying over the day.  
The third one (Model 3) is the model derived for the non-photostationary conditions in Eq. \ref{Equ-NonPhotostationary-Model}, with parameters $k_1$ and $k_3$ again varying during the day.

The models presented in sections \ref{sec:Photostationary-Theory} and \ref{sec:Chem-street-Theory} are designed to apply the chemical scheme (Eqs. \ref{Equ-Mreac1}-\ref{Equ-Mreac3}) as a post-calculation after the application of a transport and dispersion model able to provide the advected concentrations in the street, i.e. $[\ce{NO}]\!^*$, $[\ce{NO}_2]\!^*$, and $[\ce{O}_3]\!^*$.
In order to validate only the chemical models, avoiding the influence of errors due to the dispersion simulation, we have considered a virtual perfect dispersion model by using the measured concentrations in the streets as input data for the chemical models. 
According to Eqs. \ref{Equ-Photostationary-Model} and \ref{Equ-NonPhotostationary-Model}, the results of the photochemical models depend on $k_1/k_3$, $\phi_N$, $\phi_O$ and $[\ce{NO_2}]^*$. 
As mentioned above, the ratio $k_1/k_3$ can be taken as a constant or estimated from meteorological data and as a function of time.
Equations \ref{Equ-Stoechiometry-N} and \ref{Equ-Stoechiometry-O} show that the conserved quantities $\phi_N$ and $\phi_O$ can be computed directly from the measured concentrations at the monitoring stations, which correspond to $[\ce{NO}]$, $[\ce{NO}_2]$, and $[\ce{O}_3]$. 
Once $\phi_O$ is known, $[\ce{NO_2}]^*$ results from Eq. \ref{Equ-Stoechiometry-O} by subtracting the measured background concentration of ozone (see Eq. \ref{Equ-Passive-O3}).  
The other parameter of the non-photostationary model is the time scale $\tau_s$. 
In the validation, this parameter was adjusted to optimize the correlation coefficient between the model and measured concentrations. The resulting values will be discussed in the following section.

\section{On-site measurements}\label{measurements}

The field data were measured during the LYON6 campaign which took place between the 9th and the 24th July 2001 in the 6th arrondissement in Lyon (France) and was handled by COPARLY (Comité de Coordination pour le contrôle de la Pollution Atmosphérique), the local authority for traffic and air pollution management, in collaboration with the Fluid Mechanics and Acoustics Laboratory (LMFA) in \'Ecole Centrale de Lyon. 
The campaign consisted of local measurements of vehicular traffic, air pollution and weather conditions.

\begin{figure}
    \centering
    \includegraphics[width=12cm]{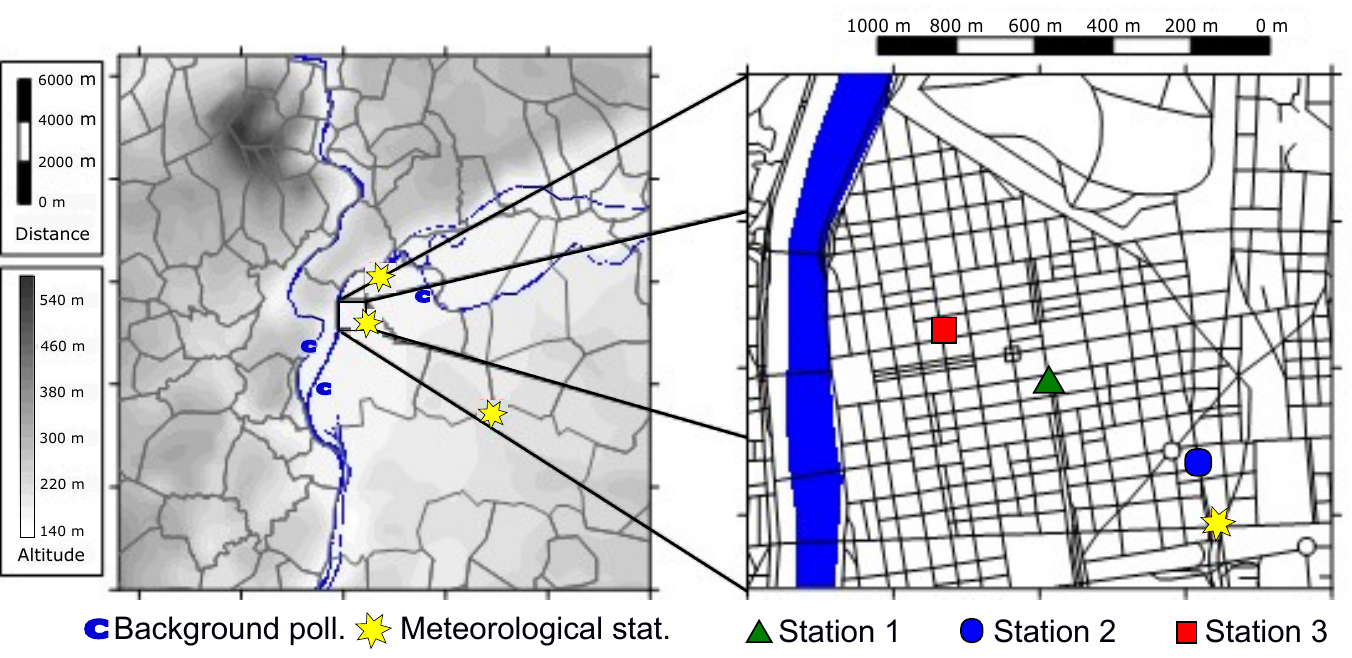}
    \caption{Location of the three meteorological stations, of the suburban stations for measuring pollution background and of the three pollution monitoring stations inside the study district.}
    \label{fig:location}
\end{figure}

\begin{figure}
    \centering
    \includegraphics[width=10cm]{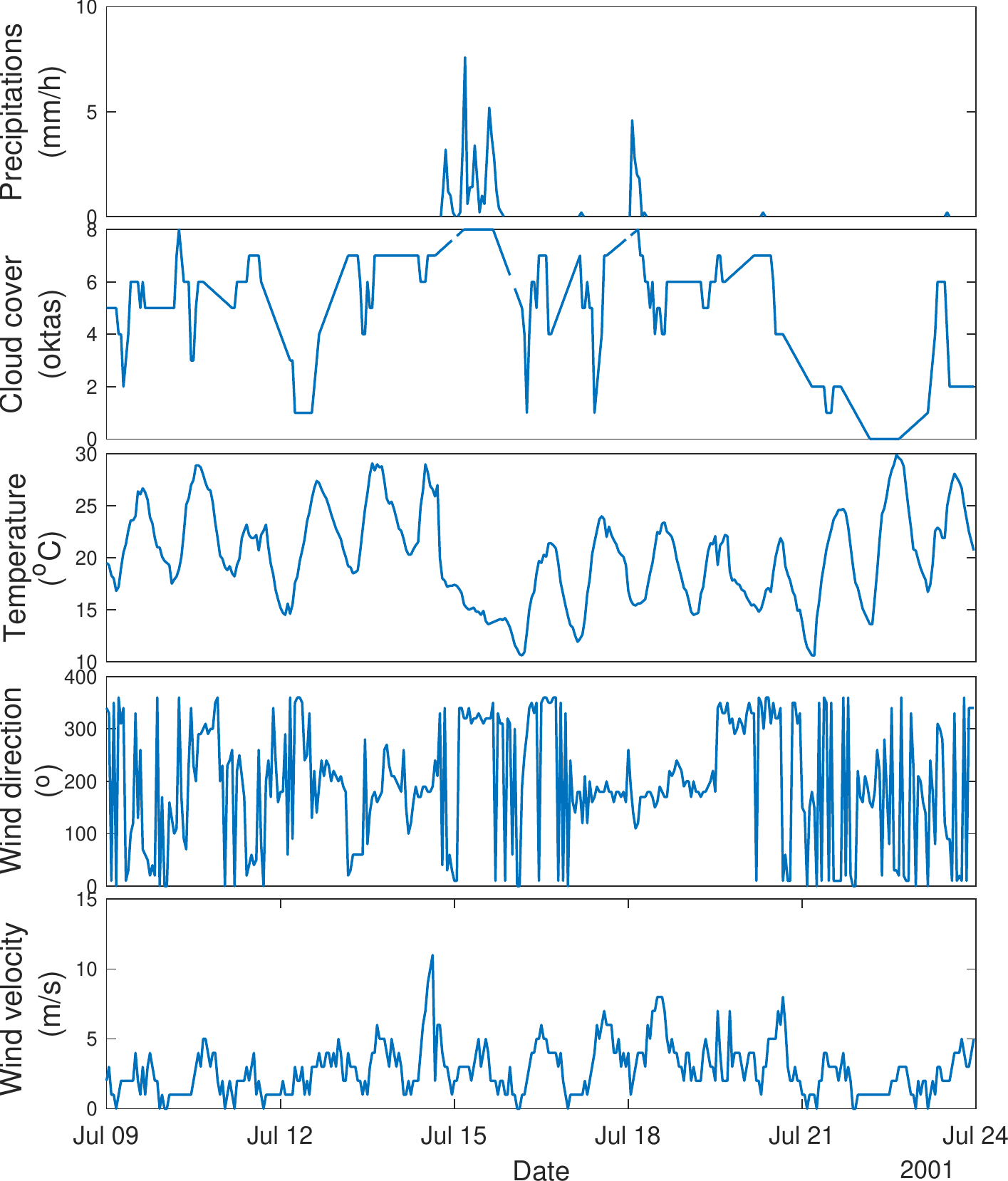}
    \caption{Temporal evolution of the meteorological parameters for the two-weeks campaign obtained by integrating the data from the different meteorological stations.}
    \label{fig:meteo}
\end{figure}

The meteorological data were collected by two stations within the urban area (Fig. \ref{fig:location}) and by a third one located 7 km from the studied district, and positioned away from any building that could directly influence the measurements.  
To have a representative dataset over the study district, the measurements from the three different stations were combined together. The reference temperature was  measured by the sensors located within the urban area. To avoid local effects, cloud cover, precipitation, wind speed and wind direction were provided by the station outside the urban area. However, a correction to the wind intensity to take into account the difference in surface roughness was applied as detailed in \citet{soulhac2012model}.
The temporal evolution of the resulting meteorological dataset is represented in Fig. \ref{fig:meteo}.

Hourly concentration of nitrogen monoxide, nitrogen dioxide and ozone were measured by three monitoring stations, referred to as `Station 1', `Station 2' and `Station 3'.  Station 1 was located in a busy street canyon. Station 2 and Station 3 were located inside school courtyards, far away from polluting source.
In this regard, we point out that the models derived in Section \ref{section_model} are valid for both street canyons and urban courtyards as the fundamental assumption underlying the box model (Eq. \ref{Equ-Adv-Diff}-\ref{Equ-Adv-Diff_O3}) is the decoupling between the dynamics in the street and the dynamics above the roofs \citep{salizzoni2011turbulent}. The urban courtyard differs from the street canyon by the absence of direct emissions and street intersections at the ends. For courtyards, therefore, the wash-out time scale $\tau_s$ (Eq. \ref{Equ-taus}) corresponds to the rate of vertical exchange at roof level ($\tau_h$) and $Q_{NO}$ and $Q_{NO_2}$ are equal to zero.

In Station 2 and 3, the analyzers were placed at 2 m from the ground and in the middle of the courtyard. The concentration was measured over approximately 6 days (see Fig. \ref{fig:meteo}).
In Station 1, the analyzer was also placed 2 m off the ground but a few centimeters from a building wall, and the concentration was measured over 15 days.
In real street canyons there are several factors (e.g., traffic, vegetation, building geometries) that increase the mixing of pollutants therefore inducing a concetration field that is more homogeneous than that observed in controlled case studies in wind tunnels and numerical simulations, where a highly spatially inhomogeneous concentration field is predicted \citep{buccolieri2009aerodynamic,marucci2019effect,fellini2020street}. For this reason, and considering that the focus of the proposed models is on hourly averaged pollution, the concentration in the canyon can be assumed sufficiently homogeneous and therefore less sensitive to the positioning of the sensor within the street. This is in line with previous validation studies of street network models \citep{soulhac2012model,soulhac2017model} in which an influence of the sensor position on the agreement between the model and the measurements was not observed.

To estimate the concentration levels of background pollutants, other three monitoring stations located outside the district were used.  
Refer to \citet{soulhac2012model} for a complete description of the measurement campaign and simulation set-up. 

\section{Results}\label{results}

\begin{figure}
    \centering
    \includegraphics[width=12cm]{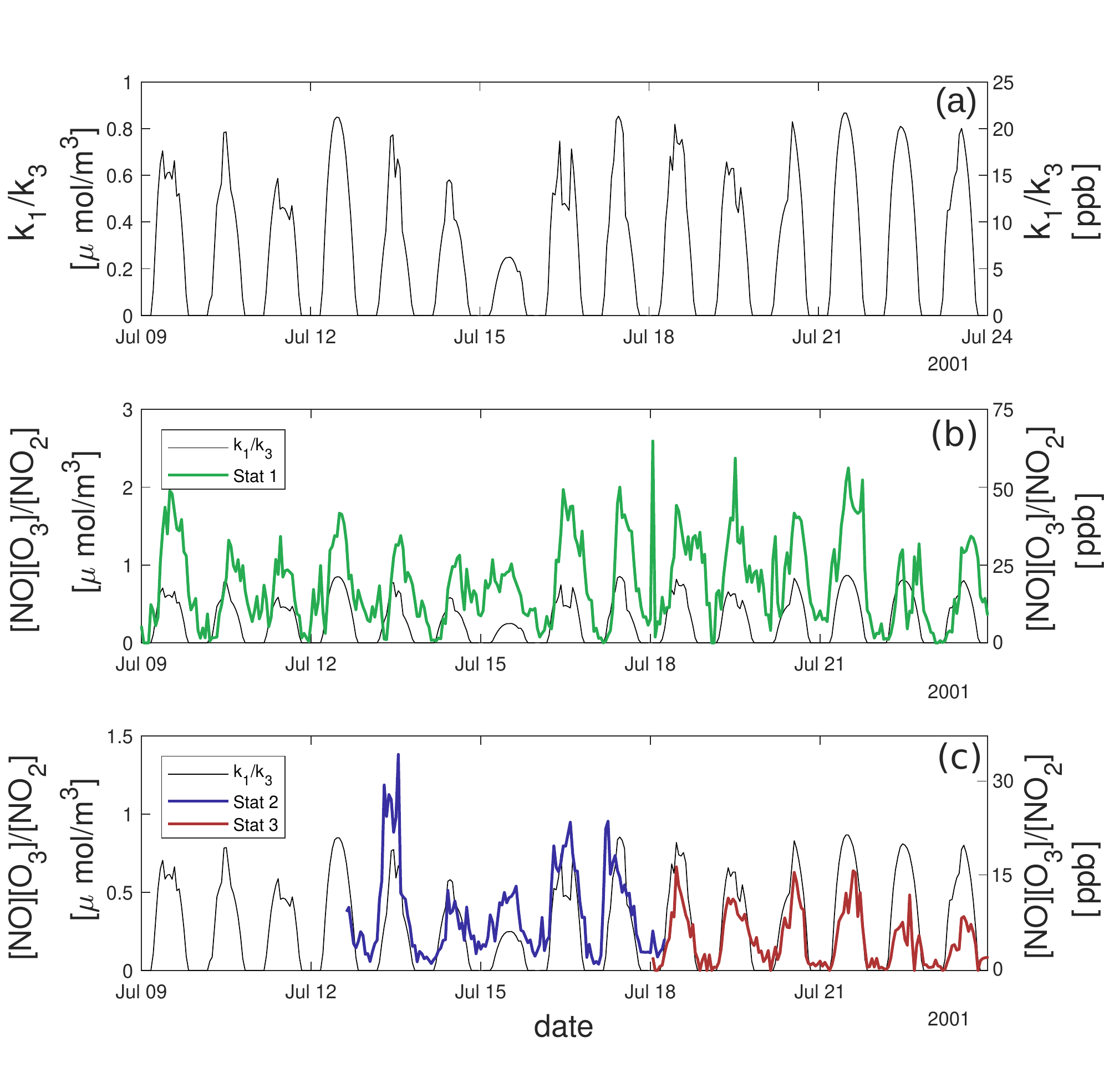}
    \caption{Evolution of the ratio $k_1/k_3$ as predicted by Eq. \ref{k1k3} (a), and comparison with the ratio $[\ce{NO}][\ce{O3}]/[\ce{NO2}]$ for the site within the street canyon -Station 1- (b) and for the two sites in the courtyards -Station 2 and Station 3- (c).}
    \label{fig:k1_k3}
\end{figure}

The meteorological data collected during the field campaign directly provide the temporal evolution of the ratio $k_1/k_3$, estimated using Eq. \ref{k1k3}. 
As shown in Fig. \ref{fig:k1_k3}-a, this ratio is far from being constant with time, and varies from a maximum of about 0.9 $\mu mol/m^3$ (20 ppb) and a minimum close to 0 during the night, with an average value of 0.3 $\mu mol/m^3$ (6.8 ppb). This is due to the variation over time of temperature, cloud coverage and solar elevation (see Eq. \ref{k1k3}). 
As stated in Eq. \ref{Equ-Photostationary-state}, this ratio equals the ratio $[\ce{NO}][\ce{O3}]/[\ce{NO2}]$ when the pollutants are in photostationary equilibrium. 
By using the measurements from the three pollution monitoring stations, we test this condition in Fig. \ref{fig:k1_k3}-b and c.
Results show that, for Station 2 and Station 3, the ratio $[\ce{NO}][\ce{O3}]/[\ce{NO2}]$ agrees well with the trend of the ratio $k_1/k_3$. This is in line with the analysis performed in Fig. \ref{fig:trend_tau_tau3}: in sites sheltered from direct vehicular emissions and with long residence times, the photostationary equilibrium (Eq. \ref{Equ-Photostationary-state}) is a reliable assumption. 
Conversely, for the busy street canyon (Station 1), the ratio $[\ce{NO}][\ce{O3}]/[\ce{NO2}]$ is generally higher than $k_1/k_3$ (Fig. \ref{fig:k1_k3}-c). This is due to the fact that, at the emission, $[\ce{NO_X}]$ are mainly constituted by $\ce{NO}$, that is progressively transformed in $\ce{NO_2}$ until photostationary equilibrium is reached. For this reason, close to the source the ratio $[\ce{NO}][\ce{O3}]/[\ce{NO2}]$ is expected to be higher than that corresponding to the equilibrium.

This first analysis shows that the photostationary model has some limitations when applied to busy street canyons with direct vehicular emissions.\\

To clarify this point, we assess the performance of the three photochemical models (derived in Section \ref{section_model} and retrieved in Section \ref{three_models}) in predicting the concentration of $[\ce{NO_2}]$ in the three measurement stations. 
For the sake of comparison, we include in our analysis the empirical model developed by \citet{dixon2001sensitivity}. Starting from the work of \citet{derwent1996empirical}, \citet{dixon2001sensitivity} developed a new \ce{NO_X}-\ce{NO_2} relationship based on a larger dataset collected across multiple sites: 
\begin{equation}
    \frac{\ce{[NO_2]}}{[\ce{NO_X}]}=A+B \log([\ce{NO_X}]) +C \log([\ce{NO_X}])^2+D \log([\ce{NO_X}])^3+E\log([\ce{NO_X}])^4.
\end{equation}
where the polynomial constants take the following values at urban sites: $A=- 3.08308$, $B=+7.472477$, $C=-5.11636$, $D=+1.381938$, $E=-0.12919$. 
For $\ce{[NO_2]}<15$ ppb, one should use $\ce{[NO_2]}/[\ce{NO_X}]= 60$ ppb. The model is reffered as DDM model in the following.\\

The four models are applied in a quasi-steady approximation, therefore describing temporal evolution of all variables (meteorological, emissions, background concentration) as the succession of stationary states lasting 1 hour.\\

The simplest photostationary model (Model 1) assumes that the ratio $k_1/k_3$ is constant over time and for the different urban locations. According to \citet{seinfeld1986atmospheric}, we assume as a typical value for this ratio 10 ppb. 
This value is in line with the time average of the trend estimated by means of Eq. \ref{k1k3} and reported in Fig. \ref{fig:k1_k3}.a.
Moreover, we have tested the sensitivity of the model results to this constant by calculating the relative variation of the mean $\ce_{NO_2}$ concentration over the simulated period for different $k_1/k_3$ ratios. 
The results are reported in Table \ref{tab:k1k3_value}.
The concentration is very sensitive to variations in the ratio $k_1/k_3$ and, as stated by Eq. \ref{Equ-Photostationary-state}, it increases as the ratio decreases. Moreover, the sensitivity of the model is higher for Station 2 and Station 3 with respect to Station 1.\\

\begin{table}[]
    \centering
\begin{tabular}{c|l|llclll}
\multicolumn{1}{l|}{}                                                                                         & $k_1/k_3$ & 2     & 5     & \multicolumn{1}{l}{10}                                                 & 15     & 20     & 25     \\ \hline
\multirow{3}{*}{\begin{tabular}[c]{@{}c@{}}Relative variation \\ of $\ce{NO}_2$\\ concentration\end{tabular}} & Station 1 & 14\% & 8\%  & \multirow{3}{*}{\begin{tabular}[c]{@{}c@{}}Ref \\ (0 \%)\end{tabular}} & -7\%  & -14\% & -20\% \\
                                                                                                              & Station 2 & 18\% & 11\% &                                                                        & -10\% & -19\% & -28\% \\
                                                                                                              & Station 3 & 20 \% & 12 \% &                                                                        & -11 \% & -21\% & -31\%
\end{tabular}
    \caption{Sensitivity of the output from Model 1 as a function of the ratio $k_1/k_3$.}
    \label{tab:k1k3_value}
\end{table}

\begin{figure}
    \centering
    \hspace*{-2cm}
    \includegraphics[width=16 cm]{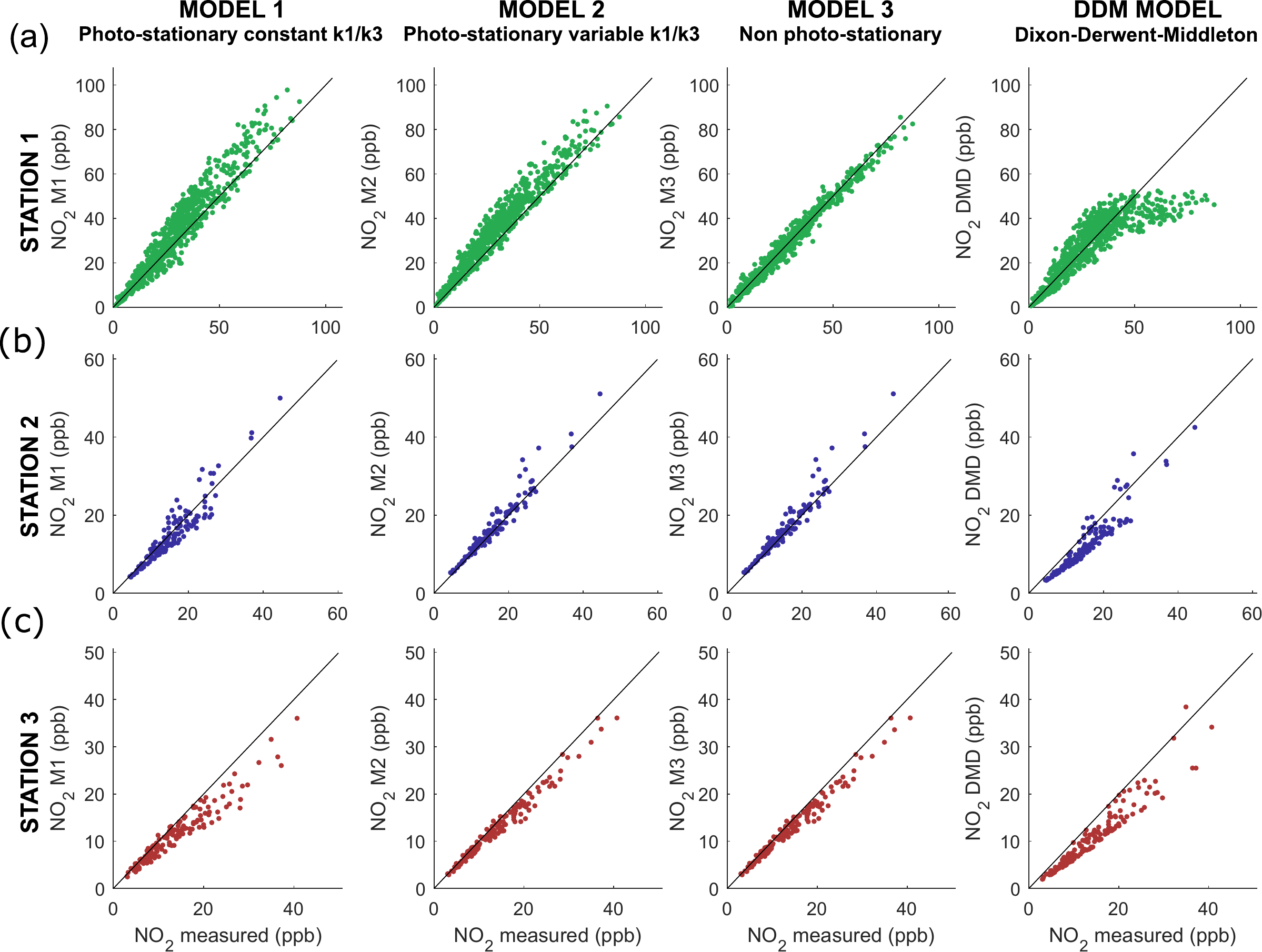}
    \caption{Comparison between the measured $\ce{NO_2}$ concentrations in Station 1 (a), Station 2 (b) and Station 3 (c) against the concentrations computed with the three different photochemical models (M1, M2 and M3) and with the Derwent-Middleton model.  Each point corresponds to one hour average concentration. Station 1 is to the busy street canyon, while Station 2 and 3 are courtyards.}
    \label{fig:scatter}
\end{figure}

The results provided by the the three photo-chemical models are shown in Fig. \ref{fig:scatter}, where measured and simulated $\ce{NO2}$ concentrations are plotted for the three monitoring stations.
Moreover, following \citet{chang2004air}, we assessed the performance of the models by means of multiple statistical indices:
\begin{itemize}
\item the Relative Error: RE$=\overline{\left( \frac{2|C_m-C_p|}{C_m+C_p} \right)}$;
\item the Fractional Bias: FB$=2(\overline{C_m}-\overline{C_p})/(\overline{C_m}+\overline{C_p})$;
\item the Normal Mean Square Error: NMSE$=\overline{(C_m-C_p)^2}/\overline{C_m} \, \overline{C_p}$;
\item the Mean Geometric bias: MG$=\exp[\overline{\ln (C_m)}-\overline{\ln(C_p)}]$;
\item the Geometrical mean squared Variance: VG$=\exp[\overline{\ln(C_m)-\ln(C_p)}^2]$;
\item the correlation coefficient:
R=$\frac{\overline{(C_m-\overline{C_m})(C_p-\overline{C_p})}}{\sigma_{Cm} \sigma_{Cp}}$;
\item the ‘fraction in a factor of 2’: fraction of the data for which $0.5 \leq C_p/C_m \leq 2$,
\end{itemize}
where $C_m$ and $C_p$ are the measured and predicted concentrations, and $\sigma_{Cm}$ and $\sigma_{Cp}$ their standard deviations. 
A perfect model would have MG, VG, R and $\mbox{FAC2}$=1, and FB, NMSE=0. 
Following \citet{chang2004air}, the performances of a dispersion model can be defined as ‘good' when the following criteria are satisfied: $|\mbox{FB}|\leq 0.3$, $\sqrt{\mbox{NMSE}}\leq 2$, $0.7\leq$ MG $\leq 1.3$, VG $\leq 1.6$, FAC2 $\geq 0.5$. In \cite{hanna2012acceptance}, the same authors suggest a relaxation of these thresholds for application in urban areas. 
While all the statistical indices are used to assess the performance of the four models, we recall that only the correlation coefficient $R$ is used as a criterion to uniquely determine the value $\tau_s$ that maximizes the correlation between the results of the non-photostationary model (Model 3) and the experimental data (see Section \ref{three_models}).\\

Panel a in Fig. \ref{fig:scatter} compares the measured and predicted concentrations for Station 1, which corresponds to the busy urban canyon with vehicular emissions.
The photostationary model with constant $k_1/k_3$ (Model 1) predicts with a good approximation the measured data but tends to overestimate $\ce{NO2}$ for mean to high concentration values.
This is confirmed by the negative fractional bias in Table \ref{tab:statistics}. 
This overestimation is also observed for low concentrations when a variable $k_1/k_3$ ratio is implemented in the photostationary model (Model 2). 
The slight loss of performance of Model 2 compared to Model 1 is highlighted by the statistical metrics in Table \ref{tab:statistics}, with the increase in the absolute value of the fractional bias and the decrease in MG from 0.91 to 0.84.
On the other hand, a noticeable improvement in the prediction is observed by applying the non-photostationary model (Model 3). 
The value of $\tau_s$ that maximizes the correlation coefficient $R$ is found to be equal to 89 s. 
This value is comparable with the time scale of the chemical reactions and thus confirms the need to adopt a non-photostationary solution (see Fig. \ref{fig:trend_tau_tau3}). 
The scatter plot in Fig. \ref{fig:scatter} shows that the dispersion of the points around the bisector decreases with respect the photostationary models, as well as the relative error (RE) in Table \ref{tab:statistics}.
Finally, the approach proposed by \citet{dixon2001sensitivity} (DDM model) fairly predicts low to medium concentrations but tends to cut the highest concentration values. The same trend was observed by \citet{vardoulakis2007operational} by applying the model by \citet{derwent1996empirical}, whose prediction is almost comparable to the DDM model for concentration up to 500 ppb. 
\citet{vardoulakis2007operational} suggested that this underestimation of the \ce{NO_2} concentration was because the empirical relationship was derived using measures that do not always reflect the typical \ce{NO_2}/\ce{NO_X} vehicle emission ratio of the case study.

Panels b and c in Fig. \ref{fig:scatter} compares the measured and predicted concentrations for Station 2 and 3, which correspond to the stations located within courtyards.
Model 1 provides slightly scattered results and noticeably underestimates \ce{NO_2} concentration in Station 3 (FB=0.18 in Table \ref{tab:statistics}). 
The adoption of a variable $k_1/k_3$ ratio (Model 2) improves the performance of the photostationary model. This is highlighted by the reduction in the relative error (RE), and by the trend towards 1 of the R metric.
On the other hand, the adoption of the non-photostationary model (Model 3) does not bring further improvements. 
For both monitoring stations, the value of $\tau_s$ that maximizes the correlation coefficient $R$ tends to infinity.
As depicted in Fig. \ref{fig:trend_tau_tau3}, this means that the results provided by the non-photostationary model ($[\ce{NO}_2]$) correspond to those provided by the photostationary one ($[\ce{NO_2}]\!^{\infty}$). 
This suggests that the pollutant concentrations in these sites already reached the photochemical equilibrium, as foreseen in Fig. \ref{fig:k1_k3}).
Finally, the approach proposed by \citet{dixon2001sensitivity}
performs worse than the three physically-based models also for Station 2 and 3. Differently from Station 1, here the predictions are significantly underestimated (FB=0.21 and 0.28) also for low to medium concentration values.

Despite differences in performance, we finally notice that the statistical metrics in Table \ref{tab:statistics} are within the validity ranges suggested by \citet{chang2004air} for all models. 

\begin{figure}
    \centering
    \hspace*{-2cm}
    \includegraphics[width=16 cm]{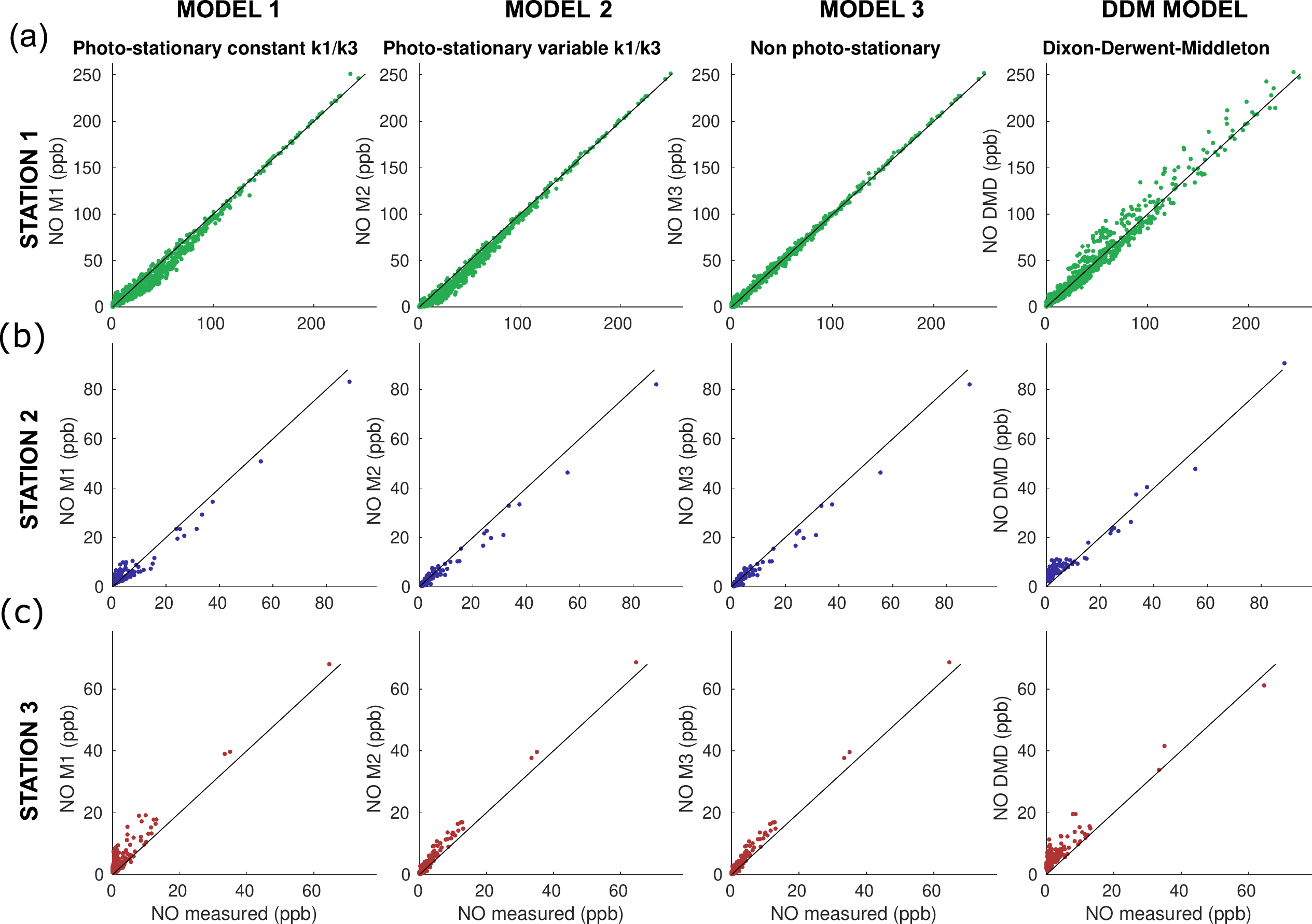}
    \caption{Comparison between the measured $\ce{NO}$ concentrations in Station 1 (a), Station 2 (b) and Station 3 (c) against the concentrations computed with the three different photochemical models (M1, M2 and M3) and with the Derwent-Middleton model.  Each point corresponds to one hour average concentration. Station 1 is to the busy street canyon, while Station 2 and 3 are courtyards.}
    \label{fig:scatterNO}
\end{figure}

In addition to \ce{NO_2} concentrations, the models derived in Section \ref{section_model} (and presented in Section \ref{three_models}) provide \ce{NO} and \ce{O_3} concentrations. 
Fig. \ref{fig:scatterNO} shows that, in a general way, the concentrations of \ce{NO} are well simulated by the proposed models. As observed  for \ce{NO_2} concentrations, the non-photostationary model (Model 3) outperforms Model 2 for Station 1, while for Stations 2 and 3 adopting the photostationary model with variable $k_1/k_3$ is sufficient to maximize correlation. 
The DDM model shows good agreement for Station 1 for high concentration values, while for medium-low values (between 50 and 100 ppb) the error is significant. In accordance with Eq. \ref{Equ-Stoechiometry-N}, this behaviour reflects the results found for \ce{NO_2} concentrations. We also observe that for stations 2 and 3, the DDM model fails to reproduce low \ce{NO} concentrations. Regarding the prediction of ozone, the same considerations made for \ce{NO} and \ce{NO_2} are valid: Model 3 brings significant improvements in results for the busy street canyon (Station 1), while the photostationary assumption (Model 2) holds when predicting concentrations in stations far from direct emissions (Station 2 and 3).

\begin{table}[]\label{tab:statistics}
\begin{tabular}{|cllllllll|}
\hline
\multicolumn{2}{|c}{}                                   & RE                           & FB                            & NMSE                         & MG                           & VG                           & R                            & FAC2                         \\ \hline
\multicolumn{1}{|c|}{}                      & Station 1 & \cellcolor[HTML]{EBF1DE}0.18 & \cellcolor[HTML]{EBF1DE}-0.12 & \cellcolor[HTML]{EBF1DE}0.05 & \cellcolor[HTML]{EBF1DE}0.91 & \cellcolor[HTML]{EBF1DE}1.01 & \cellcolor[HTML]{EBF1DE}0.96 & \cellcolor[HTML]{EBF1DE}1.00 \\
\multicolumn{1}{|c|}{}                      & Station 2 & \cellcolor[HTML]{DCE6F1}0.13 & \cellcolor[HTML]{DCE6F1}0.04  & \cellcolor[HTML]{DCE6F1}0.03 & \cellcolor[HTML]{DCE6F1}1.06 & \cellcolor[HTML]{DCE6F1}1.00 & \cellcolor[HTML]{DCE6F1}0.94 & \cellcolor[HTML]{DCE6F1}1.00 \\
\multicolumn{1}{|c|}{\multirow{-3}{*}{M1}}  & Station 3 & \cellcolor[HTML]{F2DCDB}0.17 & \cellcolor[HTML]{F2DCDB}0.18  & \cellcolor[HTML]{F2DCDB}0.07 & \cellcolor[HTML]{F2DCDB}1.18 & \cellcolor[HTML]{F2DCDB}1.03 & \cellcolor[HTML]{F2DCDB}0.96 & \cellcolor[HTML]{F2DCDB}1.00 \\ \hline
\multicolumn{1}{|c|}{}                      & Station 1 & \cellcolor[HTML]{EBF1DE}0.18 & \cellcolor[HTML]{EBF1DE}-0.15 & \cellcolor[HTML]{EBF1DE}0.04 & \cellcolor[HTML]{EBF1DE}0.84 & \cellcolor[HTML]{EBF1DE}1.03 & \cellcolor[HTML]{EBF1DE}0.97 & \cellcolor[HTML]{EBF1DE}1.00 \\
\multicolumn{1}{|c|}{}                      & Station 2 & \cellcolor[HTML]{DCE6F1}0.07 & \cellcolor[HTML]{DCE6F1}-0.05 & \cellcolor[HTML]{DCE6F1}0.02 & \cellcolor[HTML]{DCE6F1}0.95 & \cellcolor[HTML]{DCE6F1}1.00 & \cellcolor[HTML]{DCE6F1}0.97 & \cellcolor[HTML]{DCE6F1}1.00 \\
\multicolumn{1}{|c|}{\multirow{-3}{*}{M2}}  & Station 3 & \cellcolor[HTML]{F2DCDB}0.10 & \cellcolor[HTML]{F2DCDB}0.10  & \cellcolor[HTML]{F2DCDB}0.02 & \cellcolor[HTML]{F2DCDB}1.09 & \cellcolor[HTML]{F2DCDB}1.01 & \cellcolor[HTML]{F2DCDB}0.99 & \cellcolor[HTML]{F2DCDB}1.00 \\ \hline
\multicolumn{1}{|c|}{}                      & Station 1 & \cellcolor[HTML]{EBF1DE}0.09 & \cellcolor[HTML]{EBF1DE}0.00  & \cellcolor[HTML]{EBF1DE}0.01 & \cellcolor[HTML]{EBF1DE}1.01 & \cellcolor[HTML]{EBF1DE}1.00 & \cellcolor[HTML]{EBF1DE}0.99 & \cellcolor[HTML]{EBF1DE}0.99 \\
\multicolumn{1}{|c|}{}                      & Station 2 & \cellcolor[HTML]{DCE6F1}0.07 & \cellcolor[HTML]{DCE6F1}-0.05 & \cellcolor[HTML]{DCE6F1}0.02 & \cellcolor[HTML]{DCE6F1}0.95 & \cellcolor[HTML]{DCE6F1}1.00 & \cellcolor[HTML]{DCE6F1}0.97 & \cellcolor[HTML]{DCE6F1}1.00 \\
\multicolumn{1}{|c|}{\multirow{-3}{*}{M3}}  & Station 3 & \cellcolor[HTML]{F2DCDB}0.09 & \cellcolor[HTML]{F2DCDB}0.10  & \cellcolor[HTML]{F2DCDB}0.02 & \cellcolor[HTML]{F2DCDB}1.09 & \cellcolor[HTML]{F2DCDB}1.01 & \cellcolor[HTML]{F2DCDB}0.99 & \cellcolor[HTML]{F2DCDB}1.00 \\ \hline
\multicolumn{1}{|c|}{}                      & Station 1 & \cellcolor[HTML]{EBF1DE}0.21 & \cellcolor[HTML]{EBF1DE}0.05  & \cellcolor[HTML]{EBF1DE}0.09 & \cellcolor[HTML]{EBF1DE}1.04 & \cellcolor[HTML]{EBF1DE}1.00 & \cellcolor[HTML]{EBF1DE}0.86 & \cellcolor[HTML]{EBF1DE}1.00 \\
\multicolumn{1}{|c|}{}                      & Station 2 & \cellcolor[HTML]{DCE6F1}0.27 & \cellcolor[HTML]{DCE6F1}0.21  & \cellcolor[HTML]{DCE6F1}0.08 & \cellcolor[HTML]{DCE6F1}1.29 & \cellcolor[HTML]{DCE6F1}1.07 & \cellcolor[HTML]{DCE6F1}0.93 & \cellcolor[HTML]{DCE6F1}1.00 \\
\multicolumn{1}{|c|}{\multirow{-3}{*}{DDM}} & Station 3 & \cellcolor[HTML]{F2DCDB}0.32 & \cellcolor[HTML]{F2DCDB}0.28  & \cellcolor[HTML]{F2DCDB}0.12 & \cellcolor[HTML]{F2DCDB}1.38 & \cellcolor[HTML]{F2DCDB}1.11 & \cellcolor[HTML]{F2DCDB}0.96 & \cellcolor[HTML]{F2DCDB}1.00 \\ \hline
\end{tabular}
\caption{Performance statistics for the four investigated models (M1, M2, M3, DDM) in predicting the measured \ce{NO_2} concentration in the three measurement stations. Station 1 is to the busy street canyon, while Station 2 and 3 are courtyards.}
\end{table}

\begin{figure}
    \centering
    \hspace*{-0.5cm}
    \includegraphics[width=14 cm]{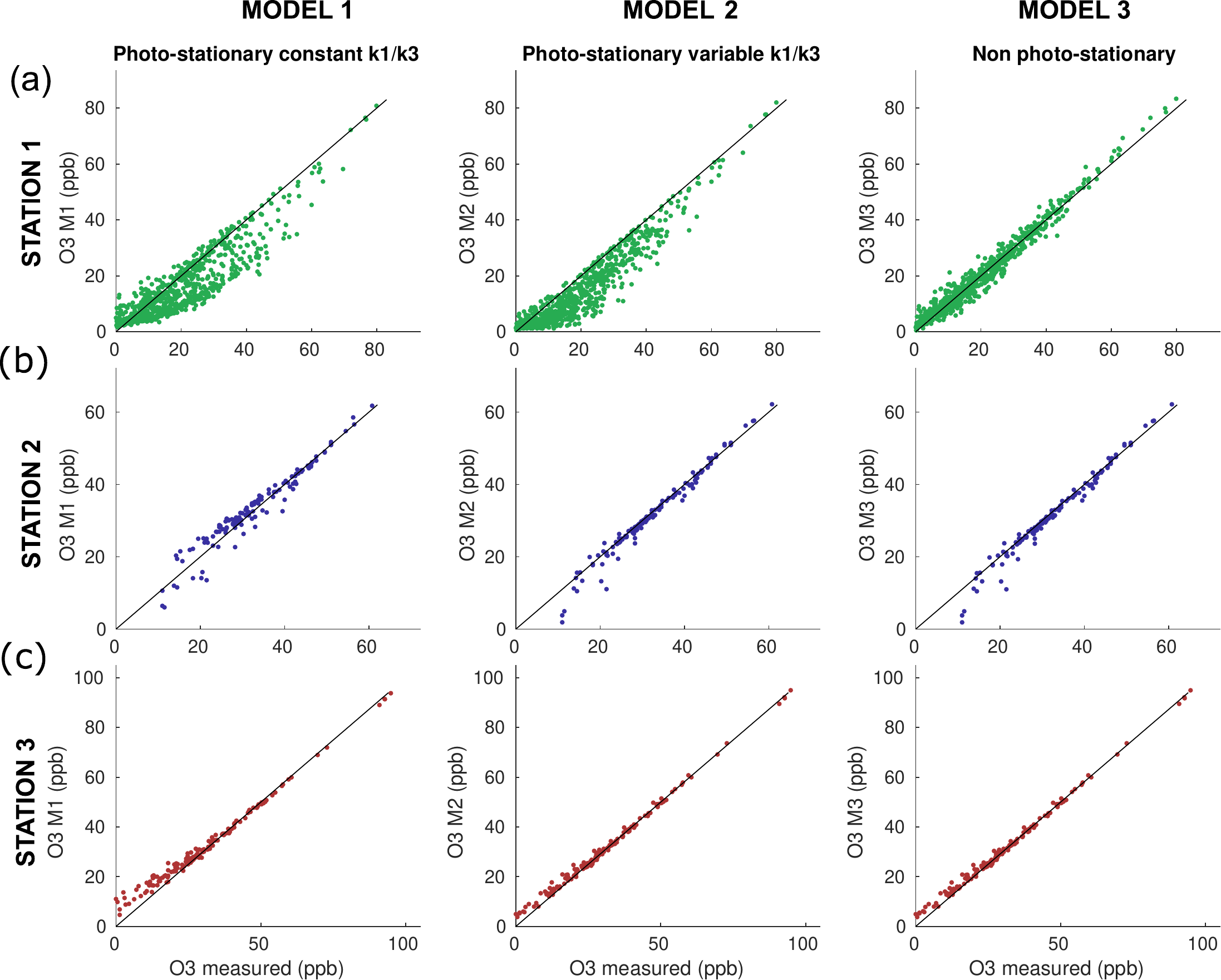}
    \caption{Comparison between the measured $\ce{O_3}$ concentrations in Station 1 (a), Station 2 (b) and Station 3 (c) against the concentrations computed with the three different photochemical models (M1, M2 and M3).  Each point corresponds to one hour average concentration. Station 1 is to the busy street canyon, while Station 2 and 3 are courtyards.}
    \label{fig:scatterO3}
\end{figure}

\section{Conclusions}\label{conclusions}

In this work we have derived different box models to simulate the concentration of $\ce{NO}$, $\ce{NO_2}$ and $\ce{O_3}$ in a street canyon. 
Starting from a mass balance in the street, we have first defined a model for a passive tracer, then for chemical species at photostationary equilibrium and finally for the non-photostationary state. 
Prediction from the simulations were compared with concentration measurements acquired during a field campaign. 
Results showed that the photostationary models adequately reproduce the pollutant concentration in canyons far from direct vehicular emissions. 
However, the implementation of a parameterization for the reaction rates according to the metereological conditions is crucial.
In busy streets, the photostationary equilibrium is not yet fully achieved and the non-photostationary model performs better.
Finally, empirical models such as Dixon-Derwent-Middleton relationship fail to reproduce concentration peaks in busy canyons and underestimate \ce{NO_2} concentrations at photochemical equilibrium.
These results show that the photostationary model with meteorology-based parameters is satisfactory in reproducing the concentrations in different urban scenarios. However, the non-photostationary model brings significant improvements in busy street canyons.

Differently from previous studies, the chemical models presented here include a description of the longitudinal and vertical ventilation processes and are therefore suitable for application to a network of streets with pollutant fluxes at street intersections. This paves the way for their implementation in operational street network models such as Sirane.

Furthermore, the adoption of a coherent formulation and the analysis of the balance equations in terms of characteristic transport and reaction times clarify the processes involved, the physico-chemical assumptions, and the limits of their validity. 
This information is critical to understanding, developing, and improving the parametric models used in existing air quality simulation software. 
In this regard, a desirable development is the treatment of non-photostationarity outside the urban canopy, i.e. over rooftops or on high-emission roads in open terrain.

Finally, we notice that the diffusion of low-cost sensors provides nowadays large databases of pollutant concentration in cities. The inclusion of more accurate transport and reaction models in operational tools for urban air pollution is in line with this growing availability of data that can be used for validation and data assimilation.
In this sense, this work highlights the feasibility of implementing non-photostationary models in simulation tools at the city scale and paves the way for further application and validation.

%\printbibliography

\end{document}